\documentstyle[prb,aps,epsfig]{revtex}
\begin{document}
\draft
\def \beq{\begin{equation}}
\def \eeq{\end{equation}}
\def \beqarr{\begin{eqnarray}}
\def \eeqarr{\end{eqnarray}}


\title{Effect of impurities on Fulde-Ferrell-Larkin-Ovchinnikov
superconductors}

\author{D. F. Agterberg}

\address{
National High Magnetic Field Laboratory and Department of Physics,
Florida State University, Tallahassee, Florida 32306 \\ and
Department of Physics, University of Wisconsin-Milwaukee, Milwaukee,
Wisconsin 53201.
}
\author{Kun Yang}
\address{
National High Magnetic Field Laboratory and Department of Physics,
Florida State University, Tallahassee, Florida 32306 }

\date{\today}
\maketitle
\begin{abstract}
We derive the Ginzburg-Landau theory of
unconventional singlet superconductors in the presence of
a Zeeman field and impurities,
to examine the resulting Fulde-Ferrell-Larkin-Ovchinnikov (FFLO) phases.
We show that the behavior of the FFLO phases in unconventional
superconductors in the presence of impurities
is qualitatively different from that  found for $s$-wave superconductors.
\end{abstract}

\pacs{74.20.De,74.25.Dw,74.80.-g}

\section{Introduction}

In 1964 Fulde and Ferrell\cite{ff}, and Larkin and
Ovchinnikov\cite{lo} demonstrated that a superconducting state
with an order parameter that oscillates spatially may be
stabilized by a large applied magnetic field or an internal
exchange field. Such a Fulde-Ferrell-Larkin-Ovchinnikov (FFLO)
state was subsequently shown to be readily destroyed by impurities
\cite{asl69} and has never been observed in conventional low-$T_c$
superconductors. The question of observing an FFLO phase in an
unconventional superconductor has only been addressed more
recently. In particular organic, heavy fermion, and high $T_c$
superconductors appear to be promising candidates for such states
\cite{gloos,yin,norman,rainer,shimahara,murthy,dupuis,modler,tachiki,geg,sh2,maki,samokhin,buzdin,yang,symington,sym2,pickett,yang00,yang01}.
These new classes of superconductors are believed to provide
conditions that are favorable to the formation of FFLO state,
because many of them are (i) strongly type II superconductors so
that the upper critical field $H_{c2}$ can easily approach the
Pauli paramagnetic limit; and (ii) layered compounds so that when
a magnetic field is applied parallel to the conducting plane, the
orbital effect is minimal, and the Zeeman effect (which is the
driving force for the formation of FFLO state) dominates the
physics. Indeed, some experimental indications of the existence of
the FFLO state have been
reported\cite{gloos,modler,geg,symington}. All of these materials
have been argued to be unconventional superconductors and in this
way differ from the case originally considered by Fulde, Ferrell,
Larkin, and Ovchinnikov. Motivated by this possibility we have
derived the Ginzburg-Landau (GL) free energy functional for
unconventional superconductors in the presence of Zeeman splitting
and impurity potential, and use it to study the possible FFLO
phases. This is not a commonly used approach of examining the FFLO
phase. To our knowledge it has previously only been discussed in
the context of clean $s$-wave superconductors by Buzdin and
Kachkachi \cite{buzdin}. However, as we show below, it represents
a very powerful approach to study the FFLO phase since the
simplicity of the resulting theory allows complexities such as non
$s$-wave pairing, impurities, and even strong-coupling effects
(though this is not done here) to be included. The stability of
the various superconducting phases and to some degree, the
topology of the superconducting phase diagram can be examined
within this approach. These results help to clarify the nature of
the FFLO phase and also can be used as a guide for a theory
extended to all temperatures and magnetic fields.

\begin{figure}
\epsfxsize=2.5 in \epsfbox{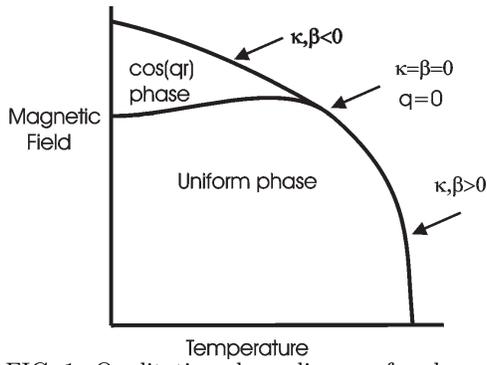} \caption{Qualitative
phase diagram for clean superconductors. The $\cos{\bf q}{\bf r}$
form of the order parameter is only valid near the normal to FFLO
transition line. The direction of ${\bf q}$ in the FFLO phase may
also depend upon temperature (see for example Ref.~[18]).}
\label{fig1}
\end{figure}

Using a functional integral formalism, we derive the GL free energy functional
for single component singlet superconductors
in the presence of impurities and Zeeman fields in the weak coupling limit.
The resultant
GL free energy is valid near the second order normal to superconductor phase transition
line in the $(T,H)$ plane. This line will be denoted by $[T,H(T)]$.
The resulting
instability to the FFLO
phase appears readily within this
approach due to the change of sign of the gradient
term $\kappa |\nabla \Psi|^2$
along the line $[T,H(T)]$.
The point at which the coefficient
$\kappa$ changes sign [denoted $[T^*,H(T^*)]$]
is a tricritical point. At this point
the normal,
uniform superconducting, and FFLO phases all meet (see Fig.~\ref{fig1}).
An intriguing feature of the weak-coupling clean limit is
that the fourth order uniform term $\beta|\Psi|^4$ also changes sign at
the tricritical point. It is this term that determines the form of the FFLO
phase. In particular,
for a given momentum ${\bf q}$ both solutions $\Psi\sim
e^{i{\bf q}\cdot{\bf r}}$
and $\Psi\sim e^{-i{\bf q}\cdot{\bf r}}$ are degenerate superconducting states
at the normal to superconductor
instability. The fourth order term breaks this degeneracy and selects
either a $\cos({\bf q}\cdot{\bf r})$ (LO phase) or a $e^{i{\bf q}\cdot{\bf r}}$ (FF phase)
type order parameter.
Since the magnitude of the order parameter is spatially uniform for the
FF phase and vanishes at lines in real space for a LO phase, a negative
$\beta$ stabilizes the LO phase. Note that the complete GL free energy
in this case requires inclusion of terms of the form
$|\Psi|^6$ and $|\Psi|^2|\nabla\Psi|^2$
to be bounded. These considerations have appeared in
the work of Buzdin and Kachkachi \cite{buzdin} for conventional
$s$-wave superconductors in the clean limit
and are shown here to remain true for unconventional superconductors.

\begin{figure}
\epsfxsize=2.5 in \epsfbox{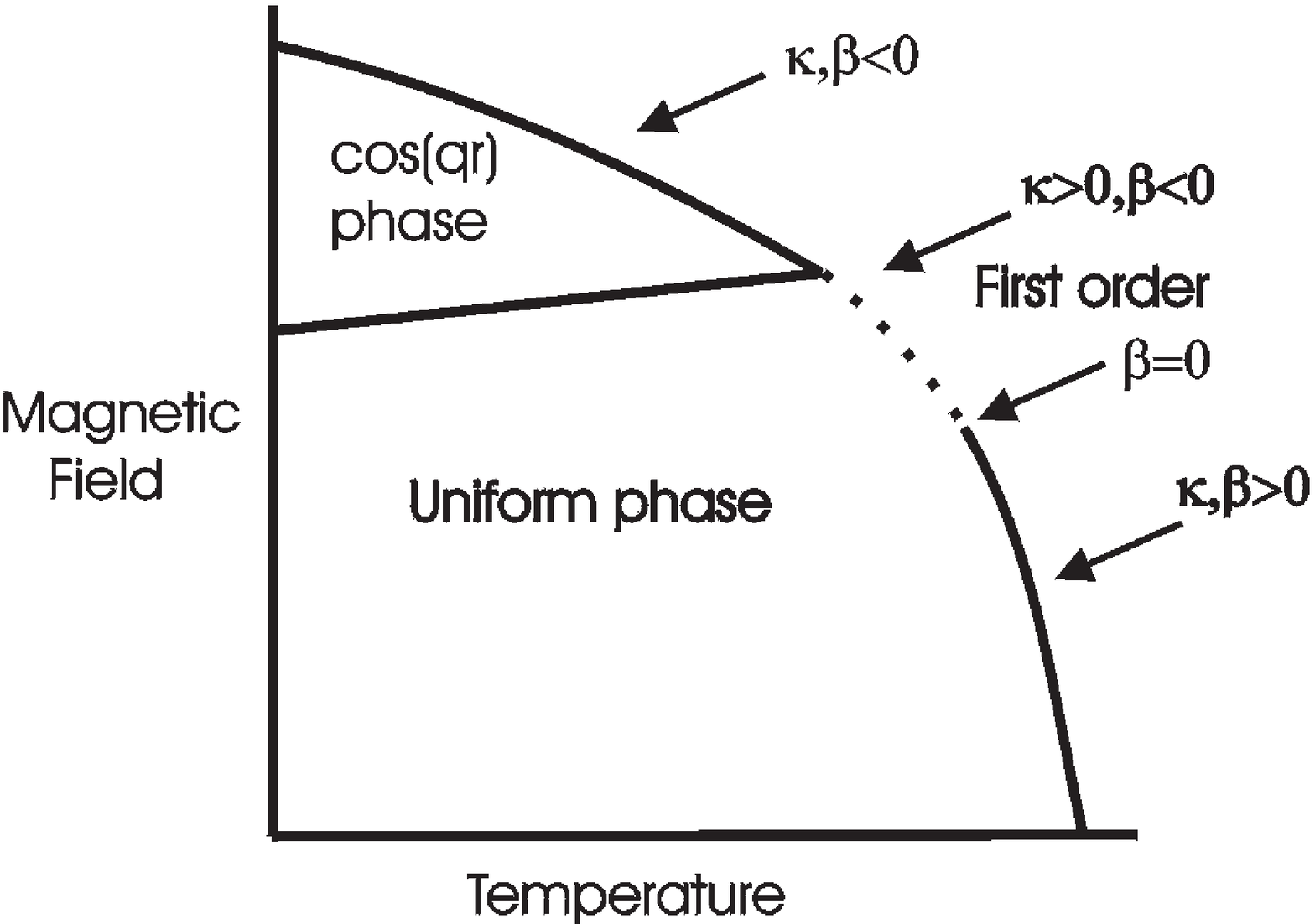} \caption{Qualitative
phase diagram for conventional superconductors with non-magnetic
impurities.} \label{fig2}
\end{figure}

\begin{figure}
\epsfxsize=2.5 in \epsfbox{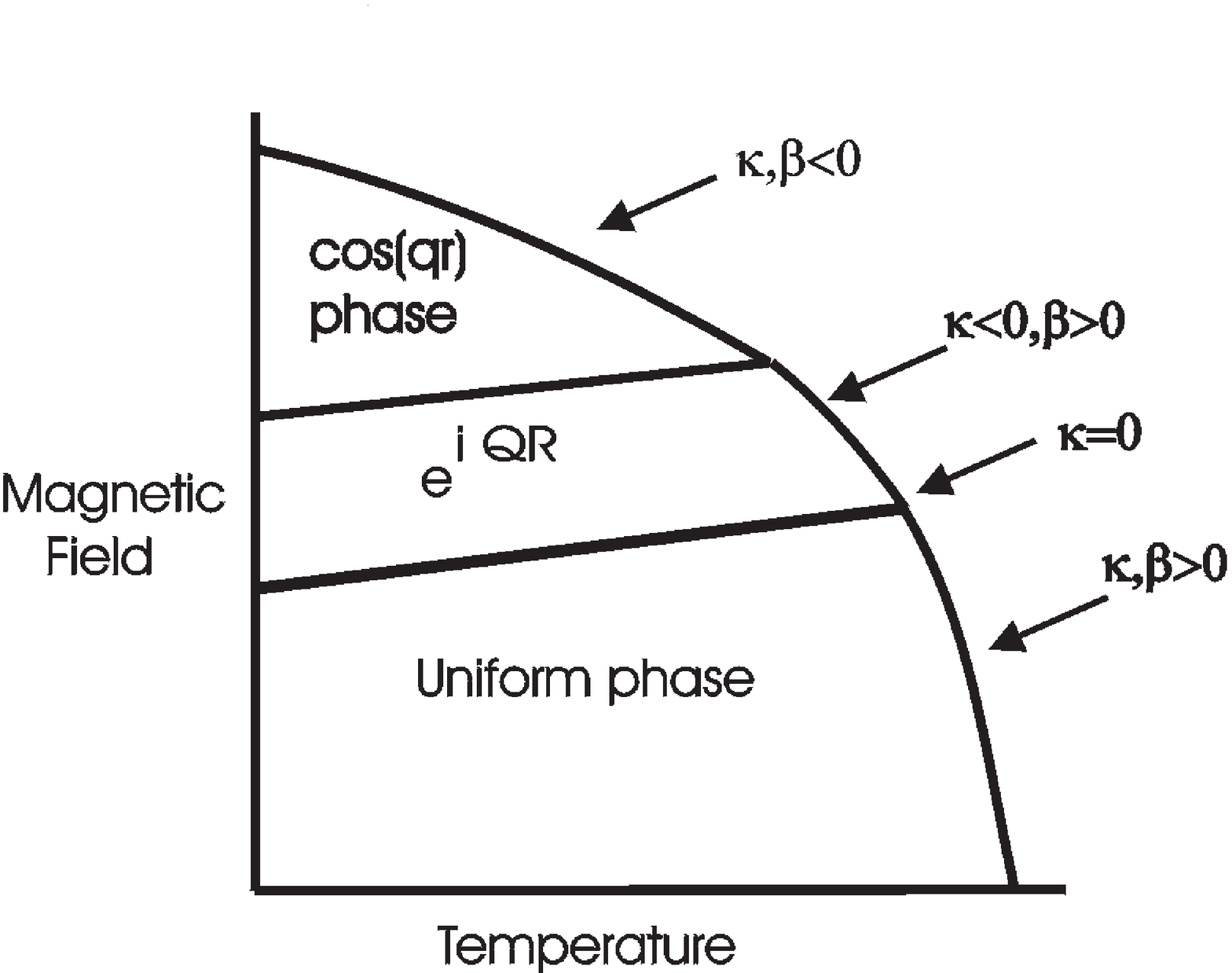} \caption{Qualitative
phase diagram for unconventional superconductors with non-magnetic
impurities.} \label{fig3}
\end{figure}

We further extend these considerations to include the effect of impurities.
It is found that impurities suppress the FFLO phase for both conventional
and unconventional superconductors.
However, we also find that impurities lead to qualitatively different
$(T,H)$ phase diagrams for conventional ($s$-wave) and
unconventional (non $s$-wave) superconductors (see Figs.~\ref{fig2} and ~\ref{fig3}).
This difference is most easily understood by looking at the coefficients
$\kappa$ and $\beta$. For conventional superconductors
it is known that
$\beta$ is unchanged by non-magnetic impurities
(this is a consequence of Anderson's
theorem) while $\kappa$ is changed. The point on the $[T,H(T)]$ line that
$\kappa$ changes sign is
pushed to lower temperatures with impurities (which illustrates that impurities
suppress the FFLO phase). Consequently,
the normal to FFLO transition in the clean limit
is replaced
by a first order normal to uniform superconducting transition.
For unconventional
superconductors impurities change
{\em both} the $\kappa$ and $\beta$ coefficients, due to the absence of
Anderson's theorem.
It is found that the points on the $[T,H(T)]$
line that $\kappa=0$ and $\beta=0$
move to lower temperatures with increasing impurity concentration.
The point $\beta=0$ is more rapidly suppressed than the point $\kappa=0$.
This implies
and that the initial instability into the FFLO phase is into a FF phase
($\Psi\sim e^{i{\bf q}\cdot{\bf r}}$)
as opposed to the LO phase ($\Psi\sim\cos{\bf q}\cdot{\bf r}$)
that is typically encountered. We are not aware of
any other report of the stability of a FF phase.
The results here also indicate that the first order
normal to uniform superconductor phase transition  does not occur for
unconventional superconductors.

\section{Ginzburg-Landau Theory}

Consider the Hamiltonian
\beq
\hat{H}=
\int{d{\bf x}}\sum_{\sigma}\Psi^{\dagger}_{\sigma}({\bf x})
[T({\bf x})+U({\bf x}) +2\sigma \mu B]
\Psi_{\sigma}({\bf x})
+\hat{V}_{int},
\eeq
where $T({\bf x})$ represents the kinetic energy and takes the form
$-\nabla^2/(2 m)-\epsilon_F$ for a free electron and more generally takes
the form $T({\bf x})=\epsilon({\bf k}=i\nabla)$ for a band with dispersion
$\epsilon({\bf k})$ measured from the Fermi energy $\epsilon_F$,
$U({\bf x})$ is the disorder potential and satisfies $\langle U({\bf x}) \rangle =0$
and $\langle U({\bf x})U({\bf x}') \rangle =n_i W\delta^d({\bf x}-{\bf x}')$, $n_i$ is
the concentration of impurities, and
$\mu=g\mu_B/2$ is the magnetic moment of the electron.
We will primarily be interested in singlet pairing in the ground state;
so we neglect
interactions between electrons with the same spin.
The pairing interaction ($\hat{V}_{int}$) is taken to have the
separable form
\beq
\hat{V}_{int}=-V_0\sum_{{\bf k},{\bf k}',{\bf q}}f_{\bf k}f^*_{{\bf k}'}
c_{{\bf k}+\frac{{\bf q}}{2},\uparrow}^{\dagger}
c_{-{\bf k}+\frac{{\bf q}}{2},\downarrow}^{\dagger}
c_{-{\bf k}'+\frac{{\bf q}}{2},\downarrow}^{\dagger}
c_{{\bf k}+\frac{{\bf q}}{2},\uparrow}^{\dagger}
\eeq
$f({\bf k})$ describes the gap dependence on the Fermi surface and is
defined to satisfy $\sum_{\bf k}|f_{\bf k}|^2=1$.
It is also understood that
a cutoff exists in momentum space so that only electrons that are close enough
to the Fermi surface interact with each other. After taking the appropriate
Fourier transforms the interaction in real space takes the form
\beq
\hat{V}_{int}=-V_0\int d{\bf x}d{\bf x}' d{\bf R} f({\bf x}) f^*({\bf x}')
\Psi_{\uparrow}^\dagger({\bf R}+\frac{{\bf x}}{2})
\Psi_{\downarrow}^\dagger({\bf R}-\frac{{\bf x}}{2})
\Psi_{\downarrow}({\bf R}-\frac{{\bf x}'}{2})
\Psi_{\uparrow}({\bf R}+\frac{{\bf x}'}{2})
\eeq
where $f({\bf x})=\frac{1}{\sqrt{V}}\sum_{\bf k}f_{\bf k}
e^{i{\bf k}\cdot{\bf x}}$,
$V$ is the volume of the system (note this definition of $f({\bf x})$ implies
$\int d{\bf x} |f({\bf x})|^2 =1$).
One may also describe the system using an Euclidean action in terms of
Grassman variables:
\beqarr
S[\Psi, \overline{\Psi}]
=& S_0[\Psi, \overline{\Psi}]\nonumber \\& -\int_0^{\beta}{d\tau}
V_0\int d{\bf x}d{\bf x}' d{\bf R} f({\bf x}) f^*({\bf x}')
\overline{\Psi}_{\uparrow}({\bf R}+\frac{{\bf x}}{2},\tau)
\overline{\Psi}_{\downarrow}({\bf R}-\frac{{\bf x}}{2},\tau)
\Psi_{\downarrow}({\bf R}-\frac{{\bf x}'}{2},\tau)
\Psi_{\uparrow}({\bf R}+\frac{{\bf x}'}{2},\tau)
\eeqarr
where $S_0$ is the action for free electrons, and $\tau$ is the imaginary time.
The partition function is
\beq
Z=\int{D\overline{\Psi}D\Psi}e^{-S[\Psi, \overline{\Psi}]}.
\eeq
We now decouple the quartic term in $S$ by introducing a pair of
Hubbard-Stratonovich
fields $\Delta({\bf R},\tau)$ and
$\overline{\Delta}({\bf R},\tau)$, which will become the
superconducting order parameter:
\beqarr
S[\Psi, \overline{\Psi}, \Delta, \overline{\Delta}]
&=&S_0-\int_0^{\beta}{d\tau}\int
{d{\bf R}d{\bf r}}[\Delta({\bf R}, \tau) f({\bf r})
\overline{\Psi}_\uparrow({\bf R}+\frac{{\bf r}}{2},\tau)
\overline{\Psi}_\downarrow({\bf R}-\frac{{\bf r}}{2},\tau)\nonumber\\
&+&\overline{\Delta}({\bf R}, \tau)f^*({\bf r})
\Psi_\downarrow({\bf R}-\frac{{\bf r}}{2},\tau)
\Psi_\uparrow({\bf R}+\frac{\bf r}{2},\tau)]
+\int_0^{\beta}{d\tau}\int
d{\bf R}{{|\Delta({\bf R}, \tau)|^2}\over
{V_0}}.
\eeqarr
With this decoupling, the fermionic action becomes quadratic, and can be
integrated out, after which we obtain an effective action in terms of the
order parameter $\Delta({\bf R}, \tau)$:
\beq
S_e[\Delta, \overline{\Delta}]
=\int_0^{\beta}{d\tau d{\bf R}}
{|\Delta({\bf R}, \tau)|^2\over
V_0}-\log Z[\Delta, \overline{\Delta}],
\label{action}
\eeq
where
\beqarr
Z[\Delta, \overline{\Delta}]
=&\int {D\overline{\Psi}D\Psi} \nonumber \\&e^{-S_0[\Psi, \overline{\Psi}]
+\int_0^{\beta} d\tau
\int
{d{\bf R}d{\bf r}}[\Delta({\bf R}, \tau) f({\bf r})
\overline{\Psi}_\uparrow({\bf R}+\frac{{\bf r}}{2},\tau)
\overline{\Psi}_\downarrow({\bf R}-\frac{{\bf r}}{2},\tau)
+\overline{\Delta}({\bf R}, \tau)f^*({\bf r})
\Psi_\downarrow({\bf R}-\frac{{\bf r}}{2},\tau)
\Psi_\uparrow({\bf R}+\frac{\bf r}{2},\tau)]}
\eeqarr

The mean-field solution corresponds to the saddle point of
$S_e[\Delta, \overline{\Delta}]$:
\beqarr
\frac{\delta S_e[\Delta, \overline{\Delta}]}{\delta\overline{\Delta}
({\bf R})}\vert_{\Delta=\Delta_s}
&={\Delta_s({\bf R})\over V_0}
-{\delta \log Z[\Delta, \overline{\Delta}]\over \delta\overline{\Delta}
({\bf R})}\vert_{\Delta=\Delta_s}\nonumber \\ &
={\Delta_s({\bf R})\over V_0}
-\int d{\bf r}f({\bf r})\langle\Psi_\downarrow({\bf R}-\frac{\bf r}{2})
\Psi_\uparrow({\bf R}+\frac{\bf r}{2})\rangle_{\Delta_s}
=0,
\label{saddle}
\eeqarr
where $\langle\rangle_{\Delta_s}$
stands for quantum and thermal averaging in the presence
of the pairing field $\Delta_s({\bf R})$. Here we have assumed a
static saddle point so that $\Delta_s$ has no $\tau$ dependence.

The functional integral formalism can be used to derive the effective
Ginzburg-Landau free energy, in the vicinity of the second order normal to uniform superconductor
transition line $[T,H(T)]$. This has been
done for the short-range attractive interactions (that gives rise to
$s$-wave pairing)\cite{popov}.
Our starting point is the effective action, Eq. (\ref{action}).
Near $[T,H(T)]$, we may make two simplifications: i) We may neglect the
$\tau$ dependence of $\Delta$ as we expect the thermal fluctuations to
dominate the quantum fluctuations;
ii) We may expand $S_e$ in powers of
$\Delta$. The quadratic terms take the form
\beq
S^{(2)}_e[\Delta, \overline{\Delta}]
=\beta\int{d{\bf R}}
{|\Delta({\bf R})|^2\over
V_0}
-\int d{\bf R} d{\bf R}' Q({\bf R}, {\bf R}')
\overline{\Delta}({\bf R})\Delta({\bf R}')
\eeq
where
\beqarr
&Q&({\bf R}, {\bf R}')=
{\delta^2\log Z\over \delta\Delta({\bf R})
\delta \overline{\Delta}({\bf R}')}\vert_{\Delta=0}\nonumber\\
&=&\int_0^\beta d\tau_1\int_0^\beta d\tau_2\int d{\bf r}d{\bf r}'
f({\bf r})f^*({\bf r}')
\langle
\Psi_\downarrow({\bf R}-\frac{\bf r}{2}, \tau_2)
\Psi_\uparrow({\bf R}+\frac{\bf r}{2}, \tau_2)
\overline{\Psi}_\uparrow({\bf R}'+\frac{{\bf r}'}{2}, \tau_1)
\overline{\Psi}_\downarrow({\bf R}'-\frac{{\bf r}'}{2}, \tau_1)\rangle_c
\nonumber\\
&=& \int_0^\beta d\tau_1\int_0^\beta d\tau_2\int d{\bf r} d{\bf r}'
f({\bf r})f^*({\bf r}')
G_{0,\downarrow}({\bf R}-\frac{\bf r}{2},{\bf R}'-\frac{{\bf r}'}{2};
\tau_2 -\tau_1)
G_{0,\uparrow}({\bf R}+\frac{\bf r}{2},{\bf R}'+\frac{{\bf r}'}{2};
\tau_2 -\tau_1)
\nonumber \\
&=&\sum_{i\omega_n}\int d{\bf r} d{\bf r}' f({\bf r})f^*({\bf r}')
G_{0,\downarrow}({\bf R}-\frac{\bf r}{2},{\bf R}'-\frac{{\bf r}'}{2};
i\omega_n)G_{0,\uparrow}({\bf R}+\frac{\bf r}{2},{\bf R}'+\frac{{\bf r}'}{2};
-i\omega_n).
\eeqarr
The quartic term takes the form
\beq
S_e^{(4)}[\Delta,\overline{\Delta}]= -\frac{1}{2}\int d{\bf R}_1
d{\bf R}_2 d{\bf R}_3 d{\bf R}_4 R({\bf R}_1,{\bf R}_2,{\bf R}_3,{\bf R}_4)
\Delta({\bf R}_1)\overline{\Delta}({\bf R}_2)\Delta({\bf R}_3)
\overline{\Delta}({\bf R}_4)
\eeq
where
\beqarr
R({\bf R}_1,{\bf R}_2,{\bf R}_3,{\bf R}_4)=& \int_0^\beta d\tau_1 d\tau_2
d\tau_3 d \tau_4
\int d{\bf r}_1 d{\bf r}_2 d{\bf r}_3 d{\bf r}_4 f({\bf r}_1)f^*({\bf r}_2)
f({\bf r}_3)f^*({\bf r}_4)\nonumber \\ &
G_{0,\uparrow}({\bf R}_4-\frac{{\bf r}_4}{2},{\bf R}_1-\frac{{\bf r}_1}{2};
\tau_4-\tau_1)
G_{0,\downarrow}({\bf R}_4+\frac{{\bf r}_4}{2},{\bf R}_3+\frac{{\bf r}_3}{2};
\tau_4-\tau_3)
\nonumber\\ &
G_{0,\uparrow}({\bf R}_2-\frac{{\bf r}_2}{2},{\bf R}_3-\frac{{\bf r}_3}{2};
\tau_2-\tau_3)
G_{0,\downarrow}({\bf R}_2+\frac{{\bf r}_2}{2},{\bf R}_1+\frac{{\bf r}_1}{2};
\tau_2-\tau_1).
\eeqarr
We will need the sixth order term as well, whose explicit expression (that
involves the product of six Green's functions)
is not included here.
The above quadratic and quartic terms apply for a particular
impurity configuration. We will average over impurity distributions
when calculating the form of the free energy. We assume
that the gap function that appears above corresponds
to the gap function averaged over impurities and that we can ignore
impurity induced correlations in the gap function and between the gap function
and the single
particle Greens functions.
To proceed further the impurity averaged
correlation functions $\langle G G\rangle$
and $\langle G G G G \rangle$ must be calculated. We determine these
within the Born approximation and much of the derivation follows that of
Werthamer
for conventional $s$-wave superconductors\cite{parks}.

The impurity averaged normal Greens functions are
\beq
\overline{G}_{0,\sigma}({\bf k};i\omega_n)=\frac{1}{i\omega_n+i\frac{1}{2\tau}sgn{\omega_n}-
\epsilon_{{\bf k}} +2\sigma\mu B}
\eeq
where $\frac{1}{2\tau}=\Gamma=\pi n_iWN(0)$
 and $\sigma$ is $1/2$ ($-1/2$)
for $\uparrow$ ($\downarrow$). Consider the average
$\overline{Q}({\bf x}_1,{\bf y}_1;{\bf x}_2,{\bf y}_2; i\omega_n)=
\langle G_{0,\downarrow}({\bf x}_1,{\bf y}_1;
i\omega_n)G_{0,\uparrow}({\bf x}_2,{\bf y}_2; -i\omega_n)\rangle
_{imp}=\overline{Q}({\bf x}_1-{\bf y}_2, {\bf y}_1-{\bf y}_2, {\bf
x}_2-{\bf y}_2; i\omega_n)$ due to translational invariance.
Summing the usual ladder diagrams shown in \cite{parks} gives the
self-consistent solution \beqarr \overline{Q}({\bf R}_1,{\bf
R}_2,{\bf R}_3;i\omega_n)= &\overline{G}_{0,\downarrow}({\bf
R}_1-{\bf R}_2;i\omega_n) \overline{G}_{0,\uparrow}({\bf
R}_3;-i\omega_n)\nonumber \\& +n_i W\int d{\bf R}
\overline{G}_{0,\downarrow}({\bf R}_1-{\bf R};i\omega_n)
\overline{Q}(0,{\bf R}_2-{\bf R},{\bf R}_3-{\bf R};i\omega_n)
\overline{G}_{0,\uparrow}({\bf R};-i\omega_n). \eeqarr This can be
solved after taking the Fourier transforms with respect to ${\bf
R}_1,{\bf R}_2$, and ${\bf R}_3$ \beqarr \overline{Q}({\bf
k}_1,{\bf k}_2,{\bf k}_3;i\omega_n)=&
\overline{G}_{0,\uparrow}({\bf k}_1;i\omega_n)
\overline{G}_{0,\downarrow}({\bf k}_2;-i\omega_n) \nonumber \\
&\left [ V \delta_{{\bf k}_1,-{\bf k}_2}+ \frac{n_i W
\overline{G}_{0,\uparrow}({\bf k}_3;i\omega_n)
\overline{G}_{0,\downarrow} ({\bf k}_1+{\bf k}_2+{\bf
k}_3;-i\omega_n)} {1-\frac{n_i W}{V}\sum_{\bf k}
\overline{G}_{0,\uparrow}({\bf k};\i\omega_n)
\overline{G}_{0,\downarrow}({\bf k}+{\bf k}_2+{\bf
k}_3;-i\omega_n)}\right ] \eeqarr Substituting this result into
Eq. 11 gives \beqarr Q({\bf R},{\bf R}')=&\frac{1}{V}\sum_{\bf
q}e^{i{\bf q}\cdot({\bf R}-{\bf R}')} \sum_{{\bf k},{\bf
k}',i\omega_n} f_{\bf k} f^*_{{\bf k}'} \\&
\overline{G}_{0,\uparrow}({\bf k}+\frac{\bf q}{2};i\omega_n)
\overline{G}_{0,\downarrow}({\bf k}'-\frac{\bf q}{2};-i\omega_n)
\left [V\delta_{{\bf k},{\bf k}'}+ \frac{n_i W
\overline{G}_{0,\uparrow}({\bf k}'+\frac{\bf q}{2};i\omega_n)
\overline{G}_{0,\downarrow}({\bf k}-\frac{\bf q}{2};-i\omega_n)}
{1-\frac{n_iW}{V}\sum_{\bf p}\overline{G}_{0,\uparrow}({\bf
p}-\frac{\bf q}{2};i\omega_n) \overline{G}_{0,\downarrow}({\bf
p}+\frac{\bf q}{2};-i\omega_n)}\right ] \nonumber \eeqarr Note
that the form of the vertex corrections found here is not the same
as for conventional $s$-wave superconductors. In particular, when
deriving the terms up to second order in the gradients, the vertex
corrections vanish for unconventional superconductors. However,
for higher order gradient terms, the vertex corrections are not
zero.

The impurity averaged correlation function that appears in $\delta
S_4$ is less straightforward to calculate. For the terms in the
free energy that are fourth order in the order parameter we
consider the only up to second order in the gradients of the order
parameter. In this case the non-zero diagrams have the same form
are those that contribute in the $s$-wave case \cite{parks}.

After performing the appropriate Taylor series expansions the following
GL free energy
for hexagonal and square lattices is found [this expression is valid
only for non $s$-wave superconductors ($\langle f_{\bf k} \rangle=0$)]
\beqarr
F=& \alpha|\Delta|^2+\beta|\Delta|^4+\kappa|\nabla\Delta|^2+\delta|\nabla^2\Delta|^2+
\mu|\Delta|^2|\nabla\Delta|^2+\eta[(\Delta^*)^2(\nabla \Delta)^2+
(\Delta)^2(\nabla \Delta^*)^2] \nonumber \\ &+\nu|\Delta|^6+
\tilde{\delta}|(\nabla_x^2-\nabla_y^2)\Delta|^2
\eeqarr
the coefficients are
\beq
\alpha=-N(0) [\ln(T_c^0/T)+\pi K_1-\pi K_1(\Gamma=0,B=0)],
\eeq
\beq
\beta=\frac{\pi N(0)}{4} (\langle |f({\bf k})|^4\rangle K_3-
\Gamma K_4),
\eeq
\beq
\kappa=\frac{\pi N(0) \langle {\bf v}_{\perp}^2({\bf k})|f({\bf k})|^2\rangle}{8}K_3,
\eeq
\beq
\delta=-\frac{\pi N(0)\langle|f({\bf k})|^2{\bf v}_{\perp}^4({\bf k})\rangle}{64}K_5,
\eeq
\beq
\mu=8\eta=-\frac{\pi N(0) \langle {\bf v}_{\perp}^2({\bf k})|f({\bf k})|^4\rangle}{4}(K_5-
\frac{\Gamma}{\langle |f({\bf k})|^4 \rangle} K_6),
\eeq
\beq
\tilde{\delta}=\frac{\pi N(0) \langle |f({\bf k})|^2
(v_x^2({\bf k})-v_y^2({\bf k}))^2\rangle}{64}(K_5+\Gamma\tilde{K}_6),
\eeq
\beq
\nu=-\frac{\pi N(0)}{8}(\langle |f({\bf k})|^6\rangle K_5-
\frac{3 \Gamma \langle |f({\bf k}|^4\rangle}{2} K_6+2\Gamma^2K_7),
\eeq
where
\beq
K_n=(2T)^{1-n}\frac{1}{\pi^n}Re\left(\sum_{\nu=0}^{\infty}\frac{1}{(\nu+z)^n}\right ),
\eeq
$z=\frac{1}{2}-i\frac{\mu B}{2\pi T}+\frac{\Gamma}{2\pi T}$, and $
\tilde{K}_6=(2T)^{-5}\frac{1}{\pi^6}Re\left[\sum_{\nu=0}^{\infty}\frac{1}{(\nu+z)^5
(\nu+\frac{1}{2}-i\frac{\mu B}{2\pi T})}\right ]$.
The $\tilde{\delta}$ term does not appear for a hexagonal lattice.
For an orthorhombic lattice the following terms also
appear in the free energy
\beqarr
\delta F=&\tilde{\kappa}(|\nabla_x\Delta|^2-|\nabla_y\Delta|^2)+
\tilde{\mu}|\Delta|^2(|\nabla_x\Delta|^2-|\nabla_y\Delta|^2) \nonumber \\ &
+\tilde{\eta}[((\nabla_x\Delta)^2-(\nabla_y\psi)^2)(\Delta^*)^2+
((\nabla_x\Delta^*)^2-(\nabla_y\Delta^*)^2)(\Delta)^2]\nonumber \\&
+\delta_2\left\{[(\nabla_x^2+\nabla_y^2)\Delta][(\nabla_x^2-\nabla_y^2)\Delta]^*
+[(\nabla_x^2+\nabla_y^2)\Delta]^*[(\nabla_x^2-\nabla_y^2)\Delta]\right\}
\eeqarr
The coefficients $\tilde{\kappa}$, $\tilde{\mu}$, and $\tilde{\eta}$ are given
by $\kappa$, $\mu$, $\eta$, with
${\bf v}_{\perp}^2$ replaced by $(v_x^2-v_y^2)$ respectively and
the coefficient $\delta_2$ is given by $\delta$ with
${\bf v}_{\perp}^4$ replaced by ${\bf v}_{\perp}^2(v_x^2-v_y^2)$.
The free energy is the main result of this paper.

\section{$s$-wave superconductors}

\begin{figure}
\epsfxsize=2.5 in
\epsfbox{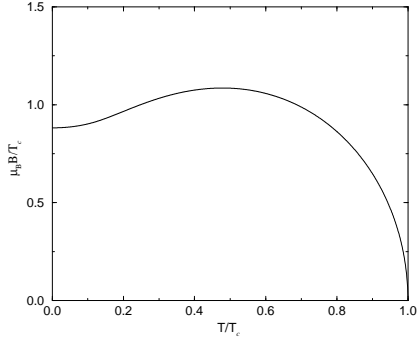}
\caption{Second order normal to uniform superconducting phase boundary
for $s$-wave superconductors. This line defines $[T,H(T)]$
when there are no impurities present. This is the boundary on which
$\kappa$ and $\beta$ are determined. Note that for $T\le0.56 T_c$
this
phase boundary will {\it not} coincide with the actual normal to
superconducting
phase boundary (the transition will either be to a non-uniform (FFLO)
phase or will be first order).}
\label{fig6}
\end{figure}

Here we review the known results about the FFLO phase for a
conventional  $s$-wave superconductor and show how it arises from
the free energy. As mentioned in the Introduction the only two
coefficients that are required to study the instability from the
normal phase into the FFLO phase near the tricritical point are
$\kappa$ and $\beta$. For a conventional $s$-wave superconductor
(for which $f_{\bf k}$ is a constant) these can easily be
determined by following Werthamer's derivation \cite{parks} (note
that the above free energy does not apply here since $\langle
f_{\bf k}\rangle \ne 0$): \beq \kappa_s=\frac{\pi N(0) \langle
{\bf v}_{\perp}^2({\bf k})\rangle}{8} \tilde{K}_3 \eeq \beq
\beta=\frac{\pi N(0)}{4}K_3(\Gamma=0) \eeq where \beq
\tilde{K}_3=(2T)^{-2}\frac{1}{\pi^3}Re\left
[\sum_{\nu=0}^{\infty}\frac{1} {(\nu+\frac{1}{2}-i\frac{\mu
B}{2\pi T})^2(\nu+\frac{1}{2}-i\frac{\mu B}{2 \pi T}
+\frac{\Gamma}{2\pi T})} \right ]. \eeq The coefficient $\beta$
does not depend upon the impurity concentration in agreement with
Anderson's theorem. The second order normal to uniform
superconductor phase line $[T,H(T)]$ is given by
$\alpha(\Gamma=0)=0$  and is shown in Fig.~\ref{fig6}. Note that
once $\kappa<0$ or $\beta<0$ the phase line $[T,H(T)]$ no longer
denotes the true normal to superconductor phase line. Numerical
evaluation of $\kappa_s$ and $\beta$ show that in the clean limit
both $\kappa$ and $\beta$ vanish at $T=0.56 T_c$. This point is
the tricritical point. The phase diagram is shown in
Fig.~\ref{fig1}. When impurities are added numerical evaluations
show $\kappa$ vanishes at a lower temperature than $\beta$
vanishes (note that the presence of impurities does not change
$\beta$). In this case there is a first order superconducting
transition to a homogeneous phase for $T\le 0.56 T_c$. Once
$\beta<0$ then the normal to superconductor instability line is no
longer given by Fig.~\ref{fig6}. To determine whether there exists
an FFLO phase for some arbitrary impurity concentration requires a
calculation that goes beyond the GL free energy presented here.
This is because the G.L. theory is only valid close to the
transition at $T=0.56 T_c$ (note that the G.L. theory can be used
to study the transition into the FFLO phase if the impurity
concentration is small enough). The calculations of Bulaevskii and
Guseinov  for layered superconductors indicate that there exists
no FFLO phase at $T=0$ when $\Gamma/T_c>0.6$ \cite{bul76}. The
qualitative phase diagram in the presence of impurities is shown
in Fig.~\ref{fig2}.

\section{Unconventional superconductors}

\begin{figure}
\epsfxsize=2.5 in \epsfbox{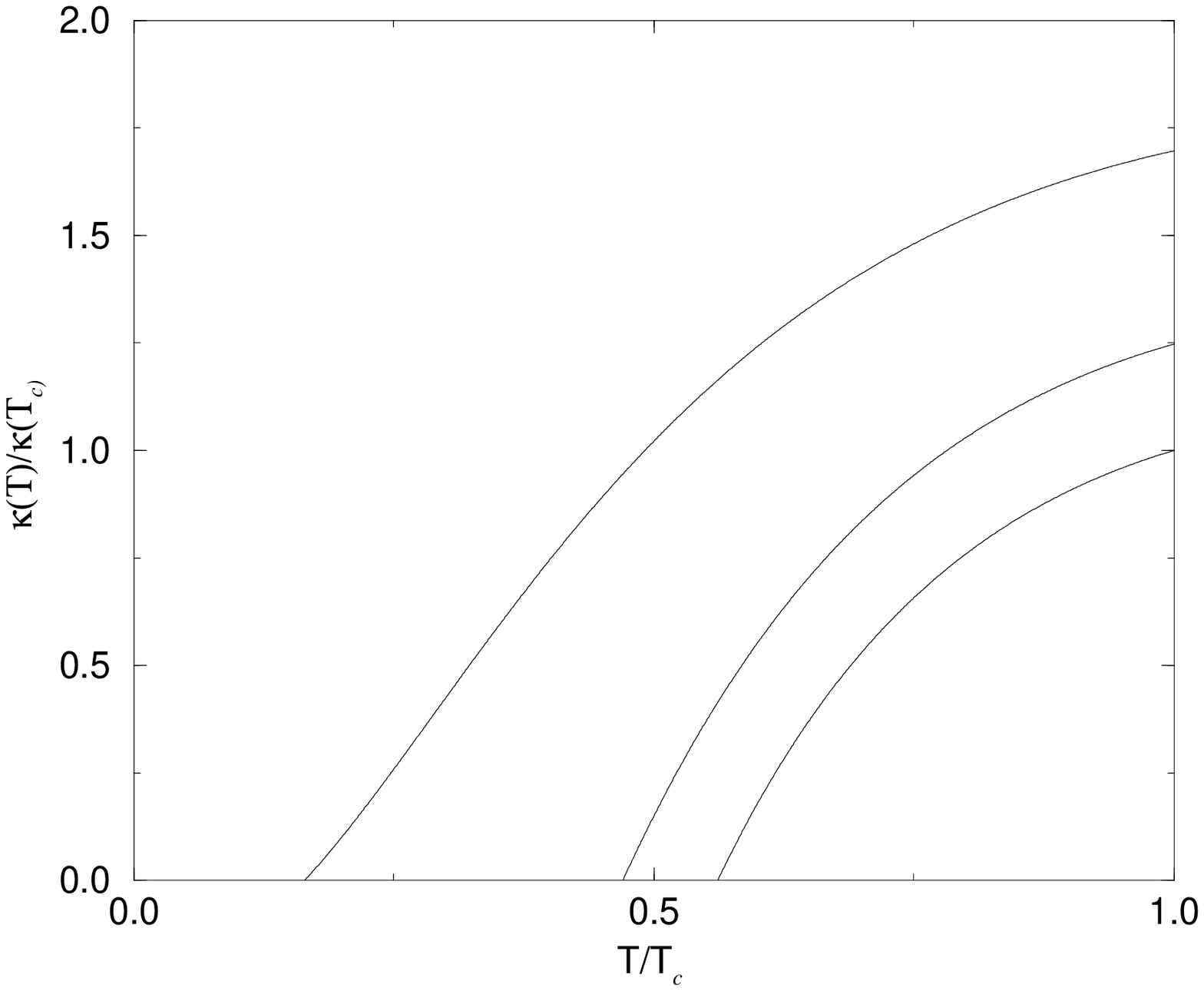}
\caption{$\kappa(T)/\kappa(T_c)$ on the phase line $[T,H(T)]$ for
$d$-wave superconductors. The curves from top to bottom at $T=T_c$
correspond to $\Gamma/T_c=0.6$, $\Gamma/T_c=0.3$, and
$\Gamma/T_c=0.0$ respectively.} \label{fig7}
\end{figure}

\begin{figure}
\epsfxsize=2.5 in \epsfbox{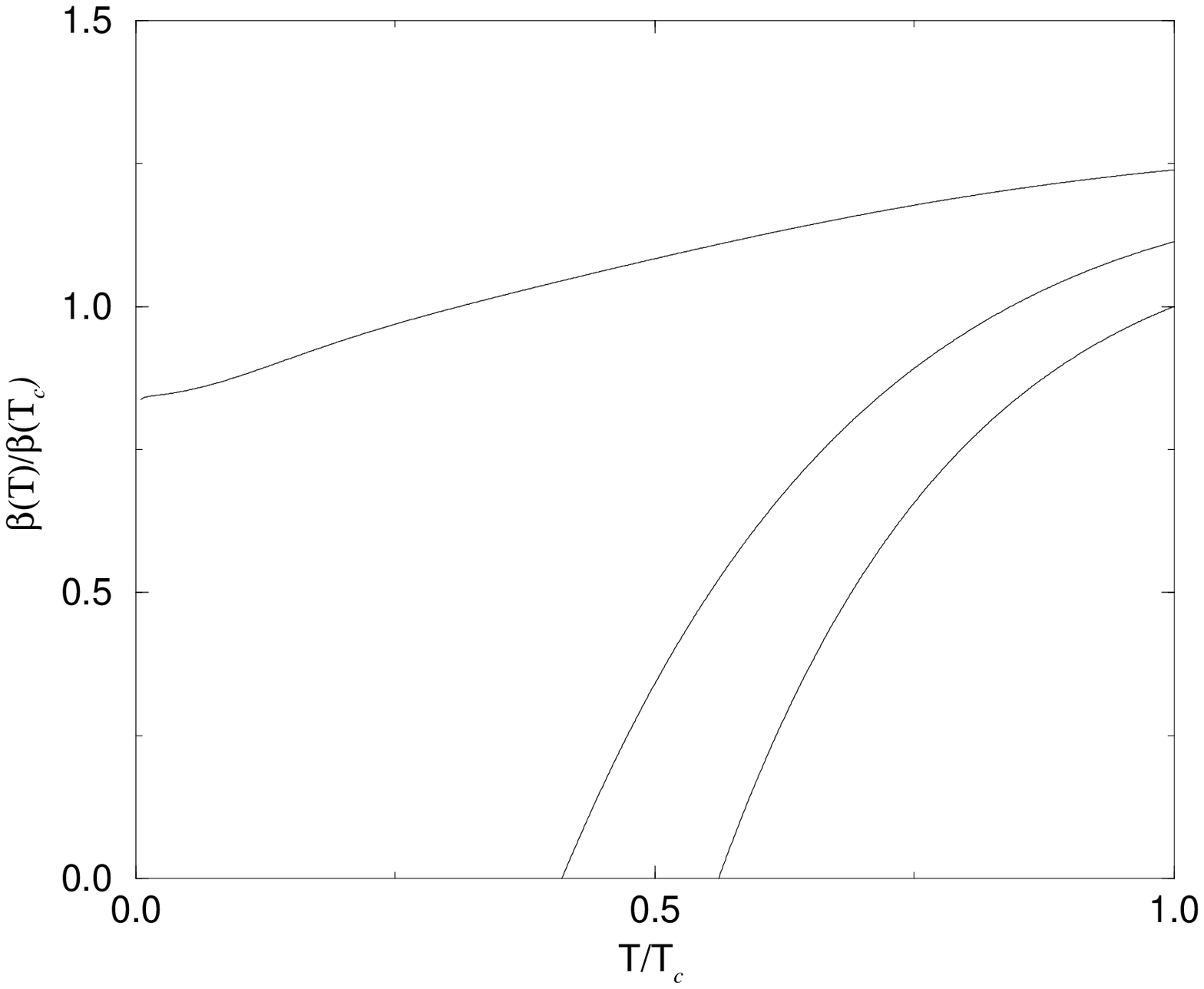}
\caption{$\beta(T)/\beta(T_c)$ on the phase line $[T,H(T)]$ for
$d$-wave superconductors. The curves from top to bottom at $T=T_c$
correspond to $\Gamma/T_c=0.6$, $\Gamma/T_c=0.3$, and
$\Gamma/T_c=0.0$ respectively.} \label{fig8}
\end{figure}

\begin{figure}
\epsfxsize=2.5 in \epsfbox{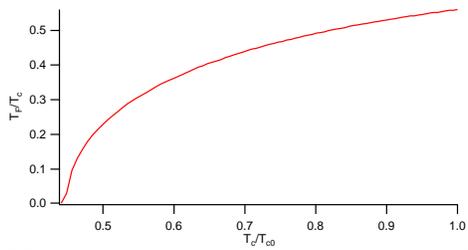} \caption{The maximum
temperature ($T_F$) for which the FFLO phase can exist as a
function of the transition temperature (which is suppressed by
impurities).} \label{fig9}
\end{figure}

The last Section demonstrated that for $T$ near $0.56T_c$ the G.L.
theory accurately reproduced the phase diagram for conventional
superconductors. Here we apply the same approach to unconventional
superconductors where it turns out the G.L. theory is more
powerful. This arises because the transition from the normal state
to the superconducting state is second order for all fields and
impurity concentrations for unconventional superconductors. For
conventional superconductors this transition is sometimes first
order, which limits the applicability of G.L. theory. For example,
the G.L. theory for unconventional superconductors can give the
maximum impurity concentration  that allows the FFLO phase to
exist; it was not able to do this for $s$-wave superconductors. As
a concrete example we study a $d$-wave superconductor ($f_{\bf
k}\propto k_x^2-k_y^2$) with a cylindrical Fermi surface. Choosing
some other $f_{\bf k}$ will not change the qualitative form of the
phase diagrams (provided $\langle f_{\bf k}\rangle=0$). The clean
limit phase diagram is qualitatively the same as that for the
$s$-wave case. In fact, the clean limit theory indicates that the
FFLO phase appears for $T<0.56T_c$ independent of the order
parameter symmetry. When impurities are added the main conclusion
is that $\kappa$ vanishes at a higher temperature than $\beta$
does when these quantities are evaluated on the phase boundary
$[T,H(T)]$ (see Fig.~\ref{fig7} and Fig.~\ref{fig8} for the
temperature evolution of $\kappa$ and $\beta$). This implies that
there is no first order transition from the normal state to a
uniform superconducting state, but rather a second order
transition into a FFLO phase. The temperature at which $\kappa=0$
gives the maximum temperature that allows the existence of the
FFLO phase (we call this temperature $T_F$). This is in sharp
contrast to what happens in $s$-wave superconductors, where a
first-order phase boundary separates the normal and uniform
superconducting state for temperatures just above the point at
which $\beta$ vanishes. Fig.~\ref{fig9} shows how $T_F$ varies as
the impurity concentration is increased (note that impurities also
suppress $T_c$, hence we plot $T_F/T_c$ as a function of
$T_c/T_{c0}$ since the latter is an experimentally measurable
quantity). This figure indicates that the FFLO phase survives a
considerable impurity concentration, only for $\Gamma/T_c\ge0.6$
does the FFLO phase cease to exist (superconductivity is destroyed
when $\Gamma/T_c\ge0.88$).

The structure of the FFLO phase can also be addressed within the G.L.
theory in the neighborhood
of tricritical point (given by $\kappa=0$ on the phase line
$[T,H(T)]$). To do this we compare the free energy for three different
phases: (1) $\Delta_1=\Delta_0$ (uniform phase), (2) $\Delta_2=\Delta_0
e^{i{\bf q}{\bf r}}$ (FF phase), and (3)$\Delta_3=\Delta_0 \cos ({\bf q}
\cdot{\bf r})$ (LO phase). We note here that terms $O(\Delta^6)$
do not need to be included in the calculations done here
because the terms $O(\Delta^4)$ are positive. Also the minimization with
respect to the orientation of ${\bf q}$ for $\Delta_2$ and $\Delta_3$
implies that ${\bf q}$ is oriented along the nodes for all impurity
concentrations. This agrees
with earlier calculations that go beyond the GL theory in the clean limit
\cite{yang,maki}.
After minimizing with respect to ${\bf q}$ the resulting free energies are
\beqarr
F_1=&\alpha|\Delta|^2+\beta|\Delta|^4\\
F_2=&(\alpha-\frac{\kappa^2}{4\delta})|\Delta|^2+
\beta_2|\Delta|^4\\
F_3=&(\alpha-\frac{\kappa^2}{4\delta})|\Delta|^2+
\beta_3|\Delta|^4\\
\eeqarr where $\beta_2=\beta-\frac{3\kappa\eta}{\delta}$ and
$\beta_3=\frac{3}{2}\beta-\frac{5\kappa\eta}{2\delta}$ (where we
have used $\mu=8\eta$). It is clear that when $\kappa=0$,
$F_1=F_2<F_3$ which implies that $\alpha=0$ and $\kappa=0$ gives
the tricritical point where the normal, uniform superconducting,
and FF superconducting phases meet. Note that the FF phase is
stable while the LO phase is not at the tricritical point since
$\beta>0$ (the FF phase is never stable in the $s$-wave case). For
temperatures below that of the tricritical point the phase
transition from the normal phase into the FFLO phase is given by
$\alpha=\frac{\kappa^2}{4\delta}$. Intriguingly, along this phase
line $\beta$ changes sign. This implies a first order transition
between the FF and the LO phase. This can be seen by comparing
$F_2$ and $F_3$ which is equivalent to comparing $\beta_2$ and
$\beta_3$. If $\beta=0$ and $\kappa<0$ then $0<\beta_3<\beta_2$
which implies $F_3<F_2$ implying that there is a phase transition
between the FF and the LO phases. For the singlet superconductors
considered here the FF phase should exhibit a spin current.

\begin{figure}
\epsfxsize=2.5 in \epsfbox{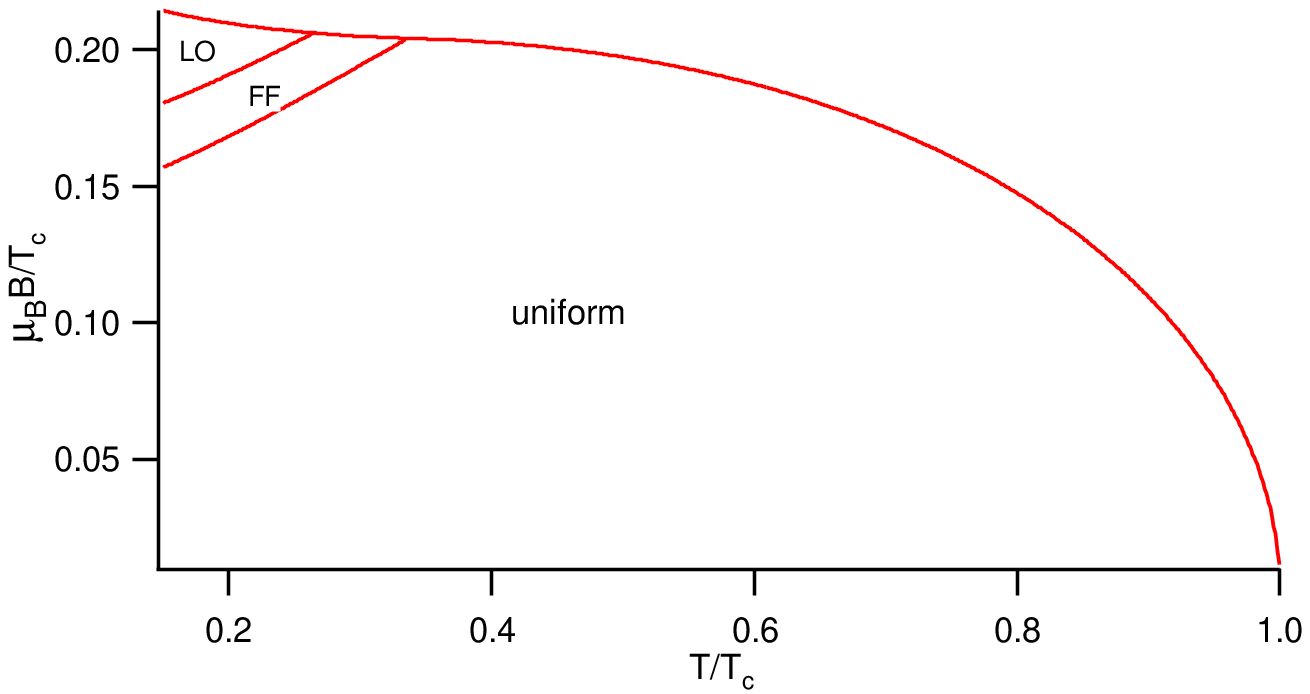} \caption {Phase diagram
for a $d$-wave superconductor with $T_c$ suppressed by impurities
such that $T_c/T_{c0}=0.573$. FF refers to an order parameter of
the form $\Delta=\Delta_0 e^{i{\bf q}{\bf r}}$ while LO refers to
an order parameter of the form $\Delta=\Delta_0 \cos( {\bf q}\cdot
{\bf r})$.} \label{fig10}
\end{figure}

A detailed calculation was carried out for an impurity
concentration for which $T_c/T_{c0}=0.573$ (where $T_{c0}$ is the
transition temperature with no impurities present).
Fig.~\ref{fig10} shows the resulting phase diagram calculated
within the G.L. theory. Note that G.L. theory in this case gives a
reasonable description of the phase diagram in this case because
it is valid along the entire normal to superconducting phase
transition boundary. For other impurity concentrations the phase
diagram is similar. If multiple scattering from the impurities
becomes important ({\it e.g.} going beyond the Born approximation
to a $T$-matrix treatment), then it is found that the region of
the phase diagram where the FF phase appears is decreased
\cite{agt02}.

\section{Conclusions}

In conclusion we have derived the GL free energy for singlet superconductors
in the presence of a Zeeman field and non-magnetic impurities.
This free energy was used to examine the resulting phase diagram.
It was shown that the phase diagrams for unconventional superconductors
and conventional superconductors are qualitatively different in the
presence of impurities. In particular the first order normal to uniform superconductor
phase transition that exists for conventional superconductors does not exist
for unconventional superconductors. Also, for unconventional superconductors
impurities induce a change in the structure of the FFLO phase. In the clean
limit the FFLO phase is described by an order parameter of  the form
$\cos({\bf q}{\bf r})$ (LO) while impurities stabilize a $e^{i{\bf q}{\bf r}}$
(FF) type order parameter.

\section{Acknowledgments}

This work was supported by NSF DMR-9971541, the Reesrach Corporation
and the A. P. Sloan Foundation
(KY), and by NSF DMR-9527035 and the State of Florida (DFA).

\begin{references}

\bibitem{ff} Fulde P and Ferrell A, Phys. Rev. {\bf 135}, A550 (1964).

\bibitem{lo} Larkin A I and  Ovchinnikov Y N, Sov. Phys. JETP {\bf 20},
762 (1965).

\bibitem{asl69} Aslamazov L G, Sov. Phys. JEPT {\bf 28}, 773 (1969).

\bibitem{gloos} Gloos K {\em et al.}, Phys. Rev. Lett. {\bf 70}, 501 (1993).

\bibitem{yin} Yin G and Maki K, Phys. Rev. B {\bf 48}, 650 (1993).

\bibitem{norman} Norman M R , Phys. Rev. Lett. {\bf 71}, 3391 (1993).

\bibitem{rainer} Burkhardt H and Rainer D, Ann. Physik {\bf 3}, 181 (1994).

\bibitem{shimahara} Shimahara H, Phys. Rev. B {\bf 50}, 12760 (1994);
J. Phys. Soc. Japan {\bf 66}, 541 (1997); {\em ibid} {\bf 67}, 736 (1998);
{\em ibid} {\bf 67}, 1872 (1998).

\bibitem{murthy}
Murthy G and Shankar R, J. Phys. Condens. Matter  {\bf 7}, 9155
(1995).

\bibitem{dupuis} Dupuis N, Phys. Rev. B {\bf 51}, 9074 (1995).

\bibitem{modler} Modler R {\em et al.}, Phys. Rev. Lett. {\bf 76}, 1292 (1996).

\bibitem{tachiki} Tachiki M {\em et al.}, Z. Phys. B {\bf 100}, 369 (1996).

\bibitem{geg} Gegebwart P {\em et al.}, Ann. Physik {\bf 5}, 307 (1996).

\bibitem{sh2} Shimahara H, Matsuo S, and Nagai K,
Phys. Rev. B {\bf 53}, 12284 (1996).

\bibitem{maki} Maki K and Won H, Czech. J. Phys. {\bf 46}, 1035 (1996).

\bibitem{samokhin} Samohkin K V, Physica C {\bf 274}, 156 (1997).

\bibitem{buzdin} Buzdin A I and Kachkachi H,
Phys. Lett. A {\bf 225}, 341 (1997).

\bibitem{yang} Yang K and Sondhi S L, Phys. Rev. B {\bf 57}, 8566 (1998).

\bibitem{symington} Symington J A {\em et al.},
preprint (2001).
\bibitem{sym2} Symington J A {\em et al.}, RWMF 2000/6th Int. Symp.
on Research in High Magnetic Fields, p. 48 (2000).

\bibitem{pickett} Pickett W E, Weht R, and Shick A B, Phys. Rev. Lett.
{\bf 83}, 3713 (1999).

\bibitem{yang00} Yang K and Agterberg D F, Phys. Rev. Lett. {\bf 84},
4970 (2000).

\bibitem{yang01} Yang K, Phys. Rev. B {\bf 63}, 140511 (2001).

\bibitem{popov} Popov V N, {\it Functional Integrals and Collective
Excitations}, Cambridge University Press, Cambridge (1987).

\bibitem{parks} Werthamer N R, in {\it Superconductivity}, edited by
R.D. Parks (Dekker, New York, 1969)

\bibitem{bul76} Bulaevskii L N  and Guseinov A A, Sov. J. Low Temp. Phys.,
{\bf 2}, 140 (1976).

\bibitem{agt02} Agterberg D F, unpublished.

\end{references}
\end{document}